\documentclass[11pt, a4paper]{article}
\usepackage[utf8]{inputenc}
\usepackage[T1]{fontenc}
\usepackage{amsmath, amssymb, amsthm}
\usepackage{graphicx}
\usepackage{booktabs}
\usepackage[margin=2.5cm]{geometry}

\newtheorem{theorem}{Theorem}[section]
\newtheorem{proposition}[theorem]{Proposition}
\newtheorem{lemma}[theorem]{Lemma}
\newtheorem{corollary}[theorem]{Corollary}
\theoremstyle{definition}
\newtheorem{definition}[theorem]{Definition}
\theoremstyle{remark}
\newtheorem{remark}[theorem]{Remark}

\newcommand{\Beta}{\mathrm{Beta}}

\title{Depth of leaves in closed linear $\lambda$-terms:\\
       exact recurrence, limit law and edge window}
\author{Olivier Bodini\thanks{LAGA, Institut Galilée, Université Sorbonne Paris Nord, CNRS, UMR 7539, F-93430 Villetaneuse, France. \texttt{olivier.bodini@univ-paris13.fr}}
}
\date{}

\begin{document}
\maketitle

\begin{abstract}
Let $K_n$ (resp.\ $D_n$) be the depth of a uniform leaf (resp.\ node) in a uniform closed linear $\lambda$-term of size $n$.
A bijective \emph{grafting identity} yields an exact polynomial-time recurrence for the full depth profile.
By the method of moments, $K_n/n$ converges in law to the density $3x/\sqrt{1-2x}$ on $[0,\frac12]$, i.e.\ $2K_n/n\to\Beta(2,\frac12)$, with $\mathbb E[K_n^r]/n^r\to3(r+1)!/(2r+3)!!$; likewise $1-2D_n/n\to\Beta(\frac12,1)$. A backward Markov chain gives the corresponding local limits for $k\le(\frac12-\varepsilon)n$.
The limit density diverges at $\frac12$, and the profile has a second scale: a window of width $\sqrt n$ around $n/2$, which carries a fraction of order $n^{-1/4}$ of the nodes, at a density $n^{1/4}$ times larger than in the bulk. We prove that a single explicit function describes both regimes: uniformly in $d$,
\[
n^{3/4}\,\mathbb P(D_n=d)=g\Bigl(\tfrac{d-n/2}{\sqrt n}\Bigr)+O(n^{-1/16}),\qquad
n^{3/4}\,\mathbb P(K_n=d)=\tfrac{3d}{n}\,g\Bigl(\tfrac{d-n/2}{\sqrt n}\Bigr)+O(n^{-1/16}),
\]
where $g(y)=2^{-1/2}e^{-y^2}D_{-1/2}(2y)$ is a parabolic cylinder function. In particular $\mathbb P(K_n>n/2)\sim\frac34\,\mathbb E|Z|^{1/2}\,n^{-1/4}\approx0.6166\,n^{-1/4}$ with $Z$ standard Gaussian. The proofs rest on harmonic functions of the heat equation evaluated along the backward chain.
\end{abstract}

\section{Introduction}
\label{sec:intro}

\paragraph{Random linear $\lambda$-terms.}
Random $\lambda$-terms are a natural model of random functional programs, and their combinatorics has attracted sustained attention. Counting closed $\lambda$-terms is already difficult: with the size taken as the number of nodes, the counting sequence grows faster than any exponential, the generating function has radius of convergence zero, and the singularity analysis of analytic combinatorics~\cite{FS09} does not apply directly~\cite{BGGJ13,GL13}. This has led to the study of restricted classes, which are interesting in their own right: terms of bounded unary height~\cite{BGG11,BGGG18}, terms in which each abstraction binds a prescribed or bounded number of variables, related to the BCI and BCK systems of combinatory logic~\cite{bodini13,BGGJ13,BG14,B26}, and terms whose size is measured on their de Bruijn representation~\cite{BGGo17}, for which a detailed statistical analysis of random terms is available~\cite{BBD19}.

Among these classes, \emph{linear} terms, in which every bound variable is used exactly once, play a special role. They are the BCI-terms, they correspond to proofs of implicative linear logic, and closed linear terms are in bijection with rooted trivalent maps~\cite{Z16}; planar variants correspond to rooted planar maps~\cite{ZG15}. Bodini, Gardy and Jacquot~\cite{bodini13} showed, through a bijection with diagrams, that there are $\sim\frac3\pi6^mm!$ closed linear terms of size $3m+2$, and gave a uniform random sampler. The limit distributions of several parameters of linear terms and of the corresponding trivalent maps, such as the numbers of closed subterms, of identity subterms and of free variables, have been determined~\cite{BSZ}, extending methods developed for random maps~\cite{BCDH18}. These works concern counts of substructures; much less is known about the \emph{geometry} of a random linear term.

\paragraph{The question.}
We study the \emph{shape} of a uniform closed linear term of size $n$, through the depth $K_n$ of a uniform leaf (variable occurrence) and the depth $D_n$ of a uniform node. For many classical families of random trees, depths live at scale $\sqrt n$ or $\log n$, and the profile has a single smooth scaling limit; for conditioned Galton--Watson trees it is the local time of the Brownian excursion~\cite{DG97}. Closed linear terms behave very differently: since every abstraction must eventually bind a variable, the terms are long and thin, and typical depths are linear in $n$.

\paragraph{Main results.}
Our starting point is a bijective \emph{grafting identity} (Theorem~\ref{thm:greffe}): removing a free variable together with its parent application yields a term with two fewer nodes and a marked node. It gives a new proof of the differential equation of~\cite{bodini13} and, more importantly, an exact recursive decomposition of closed terms, from which we obtain an exact polynomial-time recurrence for the full depth profile (Proposition~\ref{prop:recA}). We then prove the following.
\begin{enumerate}
\item \emph{Limit law} (Theorem~\ref{thm:loi}). $K_n/n$ converges in law, with all its moments, to the density $3x/\sqrt{1-2x}$ on $]0,\frac12[$, that is $2K_n/n\to\Beta(2,\frac12)$; and $1-2D_n/n\to\Beta(\frac12,1)$. In particular $\mathbb E[K_n]\sim\frac25n$ and $\mathrm{Var}(K_n)\sim\frac2{175}n^2$: depths do not concentrate.
\item \emph{Local limits in the bulk} (Theorem~\ref{thm:locale}). The mean profiles converge pointwise to these densities, uniformly on $[0,(\frac12-\varepsilon)n]$.
\item \emph{The edge window} (Theorems~\ref{thm:A}, \ref{thm:B} and~\ref{thm:masse}). The limit density diverges at $\frac12$, although depths up to $\frac23n$ are possible. We show that this singularity is smoothed in a window of width $\sqrt n$ around $n/2$, which carries a fraction of order $n^{-1/4}$ of the nodes, at a density $n^{1/4}$ times larger than in the bulk. Both regimes are captured by a single explicit formula: uniformly in $d$,
\[
n^{3/4}\,\mathbb P(D_n=d)=g\Bigl(\tfrac{d-n/2}{\sqrt n}\Bigr)+o(1),\qquad
n^{3/4}\,\mathbb P(K_n=d)=\tfrac{3d}n\,g\Bigl(\tfrac{d-n/2}{\sqrt n}\Bigr)+o(1),
\]
where $g(y)=\mathbb E[(-2V)^{-1/2};V<0]$ with $V\sim\mathcal N(y,\frac14)$, a parabolic cylinder function. Consequently $\mathbb P(K_n>n/2)\sim\frac34\,\mathbb E|Z|^{1/2}\,n^{-1/4}$ for a standard Gaussian $Z$.
\end{enumerate}
The profile of a closed linear term thus requires two renormalisations, neither of which contains the other, while a single function interpolates between them (Section~\ref{sec:deuxechelles} and Figure~\ref{fig:deuxregimes}). We are not aware of a similar two-scale profile in other natural families of random trees.

\paragraph{Methods.}
The limit law is obtained by the method of moments: the grafting decomposition gives asymptotic linear recurrences for the moments, once the contribution of terms of the form \emph{application} is controlled by a combinatorial estimate (Lemma~\ref{lem:comb}). The local limits rest on a \emph{backward Markov chain} representation of the mean profile: the mean number of nodes at depth $d$ is the expectation of a multiplicative functional of a chain that follows the ancestors of a node through the recursive decomposition. In the bulk, this chain concentrates around a deterministic path. In the edge window it does not, and we use instead explicit solutions of the backward heat equation, $h(\sigma,u)=\mathbb E\,\psi(u+\sqrt\sigma Z)$, evaluated along the chain. With $\psi(w)=|w|^{-\nu}$ they give a supermartingale controlling negative moments of the stopping size (Theorem~\ref{thm:momneg}); with $\psi(w)=|w|^{-1/2}\mathbf 1_{w<0}$ they give an approximate solution of the profile recurrence, which is the function $g$ above. An exponential supermartingale handles the tail beyond the window.

\paragraph{Organisation.}
Section~\ref{sec:exact} sets up the model, the grafting identity and the exact recurrence, and illustrates the results numerically. Section~\ref{sec:loi} proves the limit law, Section~\ref{sec:locale} the local limits in the bulk, and Section~\ref{sec:bord} the edge results. Section~\ref{sec:conclusion} concludes.

\section{The model and an exact recurrence}
\label{sec:exact}

\begin{definition}
A \emph{linear $\lambda$-term} is a $\lambda$-term (variables, abstractions, applications) in which every bound variable occurs exactly once. Terms are considered up to renaming of all variables, free ones included, so that free occurrences are anonymous holes. The \emph{size} $|t|$ is the total number of nodes of the syntax tree, leaves included. We write $\ell_{n,j}$ for the number of terms of size $n$ with $j$ free variables and $c_n=\ell_{n,0}$.
\end{definition}

\begin{proposition}
\label{prop:spec}
The series $L(z,u)=\sum_tz^{|t|}u^{\mathrm{fv}(t)}$ satisfies $L=zu+zL^2+z\,\partial_uL$.
\end{proposition}

\begin{proof}
A term is a free variable, an application (free variables add up), or an abstraction $\lambda x.t$ where, by linearity, $x$ is one of the free variables of $t$, which $\partial_u$ selects.
\end{proof}

\begin{proposition}
\label{prop:feuilles}
A term with $a$ applications, $b$ abstractions and $j$ free variables has $a+1$ leaves, size $2a+1+b$, and $j=a+1-b$. Hence a closed term has size $n=3a+2$, exactly $(n+1)/3$ leaves, and all its depths are at most $M_n:=(2n-1)/3$, this bound being attained.
\end{proposition}

\begin{proof}
Each abstraction binds a distinct leaf, so $j=(a+1)-b$; for closed terms $b=a+1$. A root-to-leaf path crosses at most the $2a+1$ internal nodes, and $\lambda x_0\cdots\lambda x_a.\,(x_0\,(x_1\,(\cdots(x_{a-1}\,x_a))))$ attains this.
\end{proof}

Since all closed terms of size $n$ have the same number of leaves, $K_n$ is the depth of a uniform (term, leaf) pair. Only sizes $n\equiv2\pmod3$ occur; $c_2,c_5,c_8,c_{11},c_{14}=1,5,60,1105,27120$.

Marking a leaf at depth $k$ by $x_k$ and writing $\mathcal S$ for the shift $x_i\mapsto x_{i+1}$, the profile series satisfies $L(z,u,\mathbf x)=ux_0z+zL(z,u,\mathcal S\mathbf x)^2+z\partial_uL(z,u,\mathcal S\mathbf x)$. Differentiating in $x_k$ at $\mathbf x=\mathbf 1$ shows that $P_k=\partial_{x_k}L|_{\mathbf x=\mathbf 1}$, which counts (term, leaf at depth $k$) pairs, satisfies $P_0=uz$ and $P_k=z(2L+\partial_u)P_{k-1}$. In coefficients, with $p_k^{(n,m)}=[z^nu^m]P_k$,
\begin{equation}
p_k^{(n,m)}=(m+1)\,p_{k-1}^{(n-1,m+1)}+2\sum_{j\le m}\ \sum_{n_1+n_2=n-1}\ell_{n_1,j}\,p_{k-1}^{(n_2,m-j)} ,
\label{eq:convo}
\end{equation}
and the closed-term profile is $p_k^{(n,0)}$. Since $c_n$ grows factorially, we work at fixed size.

\begin{theorem}
\label{thm:greffe}
$\partial_uL=z+2z^3\,\partial_zL$, i.e.\ $(j+1)\ell_{n,j+1}=2(n-2)\ell_{n-2,j}+[n=1,j=0]$.
\end{theorem}

\begin{proof}
The left side counts terms of size $n$ with $j+1$ free variables, one of which, $h$, is marked. If $n>1$, the parent $p$ of $h$ is not an abstraction (its bound variable would have to occur in the body, which is just $h$), so $p=(h\;s)$ or $(s\;h)$. Replacing $p$ by $s$ gives a linear term of size $n-2$ with $j$ free variables, a marked node (the root of $s$) and a side. Conversely, replacing the subterm $s$ rooted at a node $v$ by $(\square\;s)$ or $(s\;\square)$, with $\square$ a new marked free variable, inverts the construction.
\end{proof}

\begin{corollary}
\label{cor:EDO}
$C(z)=L(z,0)$ satisfies $C=z^2+zC^2+2z^4C'$, i.e.\ $c_n=2(n-3)c_{n-3}+\sum_{i+j=n-1}c_ic_j$ for $n\ge5$.
\end{corollary}

This equation is known~\cite{bodini13}; the bijection above gives in addition an \emph{exact decomposition}: a closed term of size $n\ge5$ is either an application of two closed terms, or $\lambda x.t$ where $t$ is obtained from a closed term $T$ of size $n-3$, a node $v$ of $T$ and a side, by grafting $(x\;s)$ or $(s\;x)$ in place of the subterm $s$ rooted at $v$. Applications form a proportion $O(1/n)$ of closed terms (Lemma~\ref{lem:comb}).

Let $A_n(h)=\sum_dA_{n,d}h^d$ (resp.\ $B_n(h)$) sum $h^{\mathrm{depth}}$ over all leaves (resp.\ nodes) of all closed terms of size $n$.

\begin{proposition}
\label{prop:recA}
$A_2=h$, $B_2=1+h$ and, for $n\ge5$ and $s=n-3$,
\begin{align*}
A_n&=2\sum_dA_{s,d}\bigl[(d+1)h^{d+2}+(s-d-1)h^{d+1}\bigr]+2h^2B_s+2h\sum_{i+j=n-1}c_jA_i,\\
B_n&=2\sum_dB_{s,d}\bigl[(d+1)h^{d+2}+(s-d-1)h^{d+1}\bigr]+2(h+h^2)B_s+2h\sum_{i+j=n-1}c_jB_i+c_n .
\end{align*}
\end{proposition}

\begin{proof}
In the grafting step above $v$ followed by the new root abstraction, a node of $T$ at depth $d$ moves to depth $d+2$ if $v$ is one of its $d+1$ ancestors-or-self, and to $d+1$ otherwise; the new application node and the new leaf sit at depths $\mathrm{depth}(v)+1$ and $\mathrm{depth}(v)+2$, the new root at $0$. In an application all depths increase by one. The factors $2$ account for the side and for the symmetry $t_1\leftrightarrow t_2$.
\end{proof}

\begin{figure}[htbp]
\centering
\includegraphics[width=\textwidth]{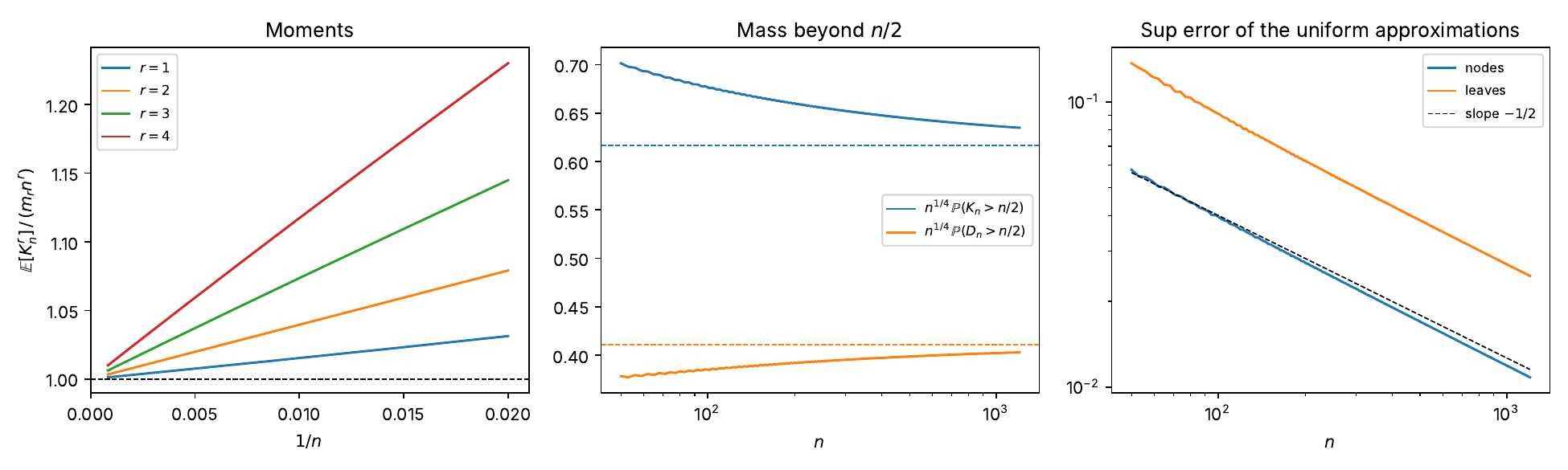}
\caption{Exact computations for $50\le n\le1202$. Left: $\mathbb E[K_n^r]/(m_rn^r)$ against $1/n$, with $m_r$ from Theorem~\ref{thm:loi}. Middle: $n^{1/4}\,\mathbb P(K_n>n/2)$ and $n^{1/4}\,\mathbb P(D_n>n/2)$ (an atom at $n/2$ is counted with weight $\frac12$), with the limits of Theorem~\ref{thm:masse} (dashed). Right: $\sup_d|n^{3/4}\mathbb P(D_n=d)-g(y)|$ and $\sup_d|n^{3/4}\mathbb P(K_n=d)-\frac{3d}ng(y)|$ (Theorems~\ref{thm:A} and~\ref{thm:B}), with a reference slope $-\frac12$.}
\label{fig:num}
\end{figure}

This recurrence agrees with~\eqref{eq:convo} for $n\le32$ and runs in exact integer arithmetic up to $n=1202$ in a couple of minutes. Figure~\ref{fig:num} uses it to illustrate the results proved below.

\section{Limit law}
\label{sec:loi}

Throughout, $n=3a+2\to\infty$, and $D_n$ is the depth of a uniform node of a uniform closed term of size $n$.

\begin{theorem}
\label{thm:loi}
For every $r\ge1$,
\[
\frac{\mathbb E[K_n^r]}{n^r}\to m_r:=\frac{3(r+1)!}{(2r+3)!!},\qquad
\frac{\mathbb E[D_n^r]}{n^r}\to\mu_r:=\frac{r!}{(2r+1)!!}.
\]
Hence $K_n/n$ converges in law to the density $\Phi(x)=3x/\sqrt{1-2x}$ on $]0,\frac12[$ (that is, $2K_n/n\to\Beta(2,\frac12)$), and $D_n/n$ to the density $(1-2x)^{-1/2}$ on $]0,\frac12[$ (that is, $1-2D_n/n\to\Beta(\frac12,1)$).
\end{theorem}

For $r\le4$ this gives $m_r=\frac25,\frac6{35},\frac8{105},\frac8{231}$ and $\mu_1=\frac13$, $\mu_2=\frac2{15}$. The limit is neither Gaussian nor $\Beta(8,12)$ (same first two moments, but third moment $\frac6{77}\ne\frac8{105}$).

\begin{remark}[Heuristic]
\label{heur:croissance}
Up to applications (a proportion $O(1/n)$ of steps), a uniform closed term grows by grafting above a uniform node and adding a root abstraction. A node at depth $d$ in a term of size $s=3t$ gains $1$, plus $1$ with probability $\approx d/(3t)$, so $d'(t)=1+d/(3t)$ and $w=\frac32-d/t$ decays as $t^{-2/3}$. Three nodes are born per step (the root with $w=\frac32$, and two at the depth of a uniform node), which yields $\frac23W\sim\Beta(\frac12,1)$ for a uniform node. This heuristic is not used in the proofs.
\end{remark}

\paragraph{Notation.}
Let $s=s_a=3a+2$, $\mathcal C_a$ the closed terms of size $s_a$, $C_a=|\mathcal C_a|$, $S_a=\sum_{b=0}^aC_bC_{a-b}$, so that $C_{a+1}=2s_aC_a+S_a$. For $f:\mathbb N\to\mathbb R$ let
\[
\beta_f(a)=\frac1{C_a}\sum_{T\in\mathcal C_a}\sum_{u\in T}f(d_T(u)),\qquad
\alpha_f(a)=\frac1{C_a}\sum_{T\in\mathcal C_a}\sum_{\ell\text{ leaf of }T}f(d_T(\ell)),
\]
$(\tau f)(d)=f(d+1)$, $B_r=\beta_{d^r}$, $A_r=\alpha_{d^r}$. Then $B_0=s$, $A_0=a+1$, $\mathbb E[D_n^r]=B_r(a)/s$, $\mathbb E[K_n^r]=A_r(a)/(a+1)$, and
\begin{equation}\label{eq:triviales}
0\le B_j(a),A_j(a)\le s^{j+1}.
\end{equation}

\begin{lemma}
\label{lem:decomp}
Let $\pi_a=2s_aC_a/C_{a+1}$ and $w_b=C_bC_{a-b}/S_a$. Then $\beta_f(a+1)=\pi_a\Lambda^{\mathrm N}_f+(1-\pi_a)\Pi^{\mathrm N}_f$ and $\alpha_f(a+1)=\pi_a\Lambda^{\mathrm F}_f+(1-\pi_a)\Pi^{\mathrm F}_f$, where, with $\mathbb E_T$ the average over $\mathcal C_a$,
\begin{align*}
\Lambda^{\mathrm N}_f&=f(0)+\tfrac1s\mathbb E_T\textstyle\sum_u\bigl[(d+1)f(d+2)+(s-d-1)f(d+1)+f(d+1)+f(d+2)\bigr],\\
\Lambda^{\mathrm F}_f&=\tfrac1s\mathbb E_T\bigl[\textstyle\sum_\ell\bigl((d+1)f(d+2)+(s-d-1)f(d+1)\bigr)+\sum_uf(d+2)\bigr],\\
\Pi^{\mathrm N}_f&=f(0)+2\textstyle\sum_bw_b\beta_{\tau f}(b),\qquad \Pi^{\mathrm F}_f=2\sum_bw_b\alpha_{\tau f}(b).
\end{align*}
\end{lemma}

\begin{proof}
Abstractions in $\mathcal C_{a+1}$ are in bijection with triples ($T$, node $v$, side), so conditionally on this form the triple is uniform; applications $(T_1\,T_2)$ have $T_1\in\mathcal C_b$, $T_2\in\mathcal C_{a-b}$ with $b\sim(w_b)$ and $T_1,T_2$ uniform. Depths move as in Proposition~\ref{prop:recA}; averaging over $v$ gives the formulas.
\end{proof}

\begin{lemma}
\label{lem:comb}
$S_a\le3C_a$ for $a\ge0$, and $M_a:=\sum_{b=1}^{a-1}C_bC_{a-b}\le3C_a/a$ for $a\ge1$. Hence $1-\pi_a\le\frac3{2s}$, $\varepsilon_a:=\sum_{b=1}^{a-1}w_b\le\frac3{2a}$ and $2w_0=2w_a=1-\varepsilon_a$.
\end{lemma}

\begin{proof}
We have $C_k\ge(6k-2)C_{k-1}$. Both inequalities are checked by exact computation for $a<30$ (in fact for $a\le35$). For $a\ge30$, assume $S_k\le3C_k$ for $k<a$, so that $C_k\le(6k+1)C_{k-1}$ for $k\le a$. With $u_b=C_bC_{a-b}$ and $2\le b\le a/2$, $u_b/u_{b-1}\le q_b:=\frac{6b+1}{6a-6b+4}$, where $q_2=\frac{13}{6a-8}$, $q_j\le\frac12$ for $j\le a/3$ and $q_j\le1$ for $j\le a/2$. Hence $u_b\le u_1q_22^{-(b-2)}$ for $b\le a/3$, and $u_b\le u_1q_22^{2-\lfloor a/3\rfloor}$ beyond; since $(a/6+1)2^{3-a/3}\le1$, $\sum_{1\le b\le a/2}u_b\le u_1(1+3q_2)=5C_{a-1}(1+\frac{39}{6a-8})$. By symmetry and $C_{a-1}\le C_a/(6a-2)$, $M_a\le\frac{10(6a+31)}{(6a-2)(6a-8)}C_a\le\frac3aC_a$ (true for $a\ge11$), and $S_a=2C_a+M_a\le3C_a$. The consequences follow from $1-\pi_a=S_a/C_{a+1}\le S_a/(2sC_a)$, $S_a\ge2C_a$ and $w_0=w_a=C_a/S_a$.
\end{proof}

\begin{proposition}
\label{prop:recasympt}
For $r\ge1$,
\begin{align*}
B_r(a+1)&=\bigl(1+\tfrac{r+2}s\bigr)B_r(a)+rB_{r-1}(a)+O(s^{r-1}),\\
A_r(a+1)&=\bigl(1+\tfrac rs\bigr)A_r(a)+rA_{r-1}(a)+\tfrac1sB_r(a)+O(s^{r-1}).
\end{align*}
\end{proposition}

\begin{proof}
Take $f(d)=d^r$. By~\eqref{eq:triviales}, the average of a sum over nodes of a polynomial of degree $\le r-2$ in $d$ is $O(s^{r-1})$. Writing $(d+1)f(d+2)+(s-d-1)f(d+1)=s(d+1)^r+(d+1)[(d+2)^r-(d+1)^r]=s\,d^r+rs\,d^{r-1}+r\,d^r+\cdots$ gives $\Lambda^{\mathrm N}_f=B_r+rB_{r-1}+\frac{r+2}sB_r+O(s^{r-1})$ and $\Lambda^{\mathrm F}_f=A_r+rA_{r-1}+\frac rsA_r+\frac1sB_r+O(s^{r-1})$. In $\Pi^{\mathrm N}_f$, the terms $b=0$ and $b=a$ give $O(1)$ and $(1-\varepsilon_a)(B_r+O(s^r))$, and the others $O(\varepsilon_as^{r+1})=O(s^r)$; so $\Pi^{\mathrm N}_f=B_r+O(s^r)$, and likewise for leaves. Finally $\beta_f(a+1)=\Lambda^{\mathrm N}_f+(1-\pi_a)(\Pi^{\mathrm N}_f-\Lambda^{\mathrm N}_f)$ with $1-\pi_a=O(1/s)$.
\end{proof}

\begin{lemma}
\label{lem:sommation}
If $x_{a+1}=(1+\frac c{a+\kappa})x_a+g_a$ with $c,\kappa>0$, $g_a\sim\gamma a^\rho$, $\gamma>0$ and $\rho>c-1$, then $x_a\sim\frac\gamma{\rho+1-c}a^{\rho+1}$.
\end{lemma}

\begin{proof}
$P_a=\prod_{k=a_0}^{a-1}(1+\frac c{k+\kappa})\sim Ka^c$, and $y_a=x_a/P_a$ satisfies $y_{a+1}-y_a=g_a/P_{a+1}\sim(\gamma/K)a^{\rho-c}$, a divergent positive series.
\end{proof}

\paragraph{Proof of Theorem~\ref{thm:loi}.}
We show $B_r\sim b_rs^{r+1}$ and $A_r\sim e_rs^{r+1}$ with $b_r=\frac{r!}{(2r+1)!!}$, $e_r=\frac{(r+1)!}{(2r+3)!!}$; for $r=0$, $b_0=1$, $e_0=\frac13$. Since $\frac{r+2}s=\frac{(r+2)/3}{a+2/3}$, Lemma~\ref{lem:sommation} with $\rho=r$ gives $B_r\sim\frac{rb_{r-1}}{2r+1}s^{r+1}=b_rs^{r+1}$, and, with $c=\frac r3$ and $g_a\sim(re_{r-1}+b_r)3^ra^r$, $A_r\sim\frac{re_{r-1}+b_r}{2r+3}s^{r+1}=e_rs^{r+1}$. Hence $\mathbb E[K_n^r]/n^r\to3e_r=m_r$ and $\mathbb E[D_n^r]/n^r\to b_r=\mu_r$. These are the moments of the announced compactly supported laws (beta integrals), which are determined by their moments; the method of moments concludes. \hfill$\square$

\section{Local limits in the bulk}
\label{sec:locale}

For $x\equiv2\pmod3$ let $\mathsf N_x(d)=x\,\mathbb P(D_x=d)$ and $\mathsf F_x(d)=\frac{x+1}3\mathbb P(K_x=d)$ be the mean numbers of nodes and leaves at depth $d$. Then $\mathsf N_x(0)=1$, $\mathsf F_x(0)=0$, both vanish outside $[0,M_x]$ and are at most $x$.

\begin{theorem}
\label{thm:locale}
For $\varepsilon\in]0,\frac12[$, uniformly in $0\le d\le(\frac12-\varepsilon)n$,
\[
\mathsf N_n(d)=\Bigl(1-\frac{2d}n\Bigr)^{-1/2}+O_\varepsilon\Bigl(\frac{\log n}{\sqrt n}\Bigr),\qquad
\mathsf F_n(d)=\frac dn\Bigl(1-\frac{2d}n\Bigr)^{-1/2}+O_\varepsilon\Bigl(\frac{\log n}{\sqrt n}\Bigr).
\]
Equivalently $n\mathbb P(K_n=k)=\Phi(k/n)+O_\varepsilon(\log n/\sqrt n)$ and $n\mathbb P(D_n=k)=(1-2k/n)^{-1/2}+O_\varepsilon(\log n/\sqrt n)$.
\end{theorem}

We give a self-contained proof of the node statement by a backward-chain argument based on Lemma~\ref{lem:comb}. The leaf statement is new.

\paragraph{Exact recurrences.}
Let $a\ge4$, $x=s_{a+1}$, $s=x-3$, $\pi=\pi_a$ and $m_a=\sum_{b=2}^{a-2}w_b$. For $1\le d\le M_x$ set
\begin{align*}
\alpha^{\mathsf N}_x(d)&=\pi\tfrac{s-d+1}s+2(1-\pi)w_a, &\beta^{\mathsf N}_x(d)&=\pi\tfrac ds,&
\alpha^{\mathsf F}_x(d)&=\pi\tfrac{s-d}s+2(1-\pi)w_a, &\beta^{\mathsf F}_x(d)&=\pi\tfrac{d-1}s,
\end{align*}
$\gamma_x=2(1-\pi)w_{a-1}$ and $W^\bullet_x=\alpha^\bullet_x+\beta^\bullet_x+\gamma_x$, which does not depend on $d$.

\begin{lemma}
\label{lem:recloc}
For $a\ge4$ and $1\le d\le M_x$, with $\mathsf V\in\{\mathsf N,\mathsf F\}$,
\[
\mathsf V_x(d)=\alpha^{\mathsf V}_x(d)\mathsf V_s(d-1)+\beta^{\mathsf V}_x(d)\mathsf V_s(d-2)+\gamma_x\mathsf V_{x-6}(d-1)+\Sigma^{\mathsf V}_x(d),
\]
where
\begin{align*}
\Sigma^{\mathsf N}_x(d)&=2(1-\pi)\bigl[w_0(\mathbf 1_{d=1}+\mathbf 1_{d=2})+w_1\mathsf N_5(d-1)+\textstyle\sum_{b=2}^{a-2}w_b\mathsf N_{s_b}(d-1)\bigr],\\
\Sigma^{\mathsf F}_x(d)&=\pi\,\mathsf N_s(d-2)/s+2(1-\pi)\bigl[w_0\mathbf 1_{d=2}+w_1\mathsf F_5(d-1)+\textstyle\sum_{b=2}^{a-2}w_b\mathsf F_{s_b}(d-1)\bigr].
\end{align*}
\end{lemma}

\begin{proof}
Apply Lemma~\ref{lem:decomp} to $f=\mathbf 1_{\{\cdot=d\}}$. In the grafting part, an old node at depth $d-2$ reaches $d$ with weight $(d-1)/s$, plus $1/s$ for the new node two levels below $v$; an old node at depth $d-1$ reaches $d$ with weight $(s-d)/s$, plus $1/s$ for the new application node. Old leaves behave alike, and the new leaf at depth $d_T(v)+2$ gives $\mathsf N_s(d-2)/s$. In the application part we separate $b=a,a-1,0,1$ and $2\le b\le a-2$, using $w_0=w_a$, $w_1=w_{a-1}$.
\end{proof}

\begin{lemma}
\label{lem:coefloc}
There is $C$ such that for $a\ge4$ and $1\le d\le M_x$:
(1) $1-\pi\le C/x$, $w_1\le C/x$, $m_a\le C/x^2$;
(2) $|W^{\mathsf N}_x-(1+\frac1x)|\le C/x^2$ and $|W^{\mathsf F}_x-(1-\frac1x)|\le C/x^2$;
(3) $|\beta^\bullet_x(d)/W^\bullet_x-d/x|\le C/x$ and $\gamma_x/W^\bullet_x\le C/x^2$;
(4) $0\le\Sigma^{\mathsf N}_x(d)\le C(x^{-1}\mathbf 1_{d\le2}+x^{-2})$ and $0\le\Sigma^{\mathsf F}_x(d)-\pi\mathsf N_s(d-2)/s\le C(x^{-1}\mathbf 1_{d\le2}+x^{-2})$.
\end{lemma}

\begin{proof}
(1) follows from Lemma~\ref{lem:comb} and its proof: $w_1\le5C_{a-1}/(2C_a)$ and, for $a\ge30$, $m_a\le195/((6a-8)(6a-2))$. Since $2w_a+2w_{a-1}=1-m_a$, $W^{\mathsf N}=\pi(1+\frac1s)+(1-\pi)(1-m_a)$ and $W^{\mathsf F}=\pi(1-\frac1s)+(1-\pi)(1-m_a)$, giving (2); (3) follows. In (4), the $w_0$ terms have weight $\le C/x$ and live on $d\le2$, the $w_1$ terms have weight $\le C/x^2$, and the central sum is at most $2(1-\pi)m_ax\le C/x^2$.
\end{proof}

\paragraph{The backward chain.}
Fix $\bullet\in\{\mathsf N,\mathsf F\}$, a size $n$ and $1\le d_0\le M_n$. The \emph{backward chain} $(X_k,H_k)$ starts at $(n,d_0)$; from an \emph{alive} state $(x,d)$, i.e.\ $x\ge17$ and $1\le d\le M_x$, it moves to $(x-3,d-1)$, $(x-3,d-2)$, $(x-6,d-1)$ with probabilities $\alpha^\bullet_x(d)/W^\bullet_x$, $\beta^\bullet_x(d)/W^\bullet_x$, $\gamma_x/W^\bullet_x$, and it stops at the first time $\tau$ its state is not alive. Let
\[
\mathcal W_k=\prod_{i<k}W^\bullet_{X_i},\qquad Z^\bullet=\sum_{k<\tau}\mathcal W_k\Sigma^\bullet_{X_k}(H_k)+\mathcal W_\tau\mathsf V_{X_\tau}(H_\tau),
\]
with $\mathsf V=\mathsf N$ or $\mathsf F$ according to $\bullet$.

\begin{lemma}
\label{lem:FK}
$\mathsf V_n(d_0)=\mathbb E[Z^\bullet]$. Moreover $\mathcal W^{\mathsf N}_k\le C(n/X_k)^{1/3}$ and $\mathcal W^{\mathsf F}_k\le C$.
\end{lemma}

\begin{proof}
The identity follows by induction on $n$ from Lemma~\ref{lem:recloc}; stopped states carry their exact values ($0$ beyond $M_X$ or below $0$, $\mathsf N(0)=1$, $\mathsf F(0)=0$, at most $14$ if $X<17$). The visited sizes are distinct, so Lemma~\ref{lem:coefloc}(2) gives the bounds on $\mathcal W_k$.
\end{proof}

\paragraph{Concentration.}
Fix $\varepsilon$ and $d_0\le(\frac12-\varepsilon)n$; let $\delta=d_0/n$, $x^*=(1-2\delta)^{3/2}n$ and $\chi(x)=\frac x2-(\frac12-\delta)n^{2/3}x^{1/3}$. Then $\chi(n)=d_0$, $\chi(x^*)=0$, $3\chi'(x)=1+\chi(x)/x$, $\chi'>\frac15$ on $[x^*/2,\infty[$, $0\le\chi''\le C_\varepsilon/n$ there, and
\begin{equation}\label{eq:chi}
\tfrac12-\chi(x)/x=(\tfrac12-\delta)(n/x)^{2/3}\ge\varepsilon .
\end{equation}
Let $\hat x_k=n-3k$, $\hat k_1$ the first $k$ with $\hat x_k<x^*/2$, $G$ the event that no step of size $6$ occurs before $\tau\wedge\hat k_1$, and $E=\{\max_{k\le\tau\wedge\hat k_1}|H_k-\chi(\hat x_k)|\le\sqrt n\log n\}$.

\begin{lemma}
\label{lem:conc}
For $n\ge n_\varepsilon$: $\mathbb P(G^c)\le C_\varepsilon/n$ and $\mathbb P(G\cap E^c)\le2e^{-c_\varepsilon(\log n)^2}$. On $G\cap E$: $\tau\le\hat k_1$, $X_k=\hat x_k$ for $k\le\tau$, $H_\tau\in\{0,-1\}$, $|X_\tau-x^*|\le6\sqrt n\log n$, and $H_k/X_k\le\frac12-\frac\varepsilon2$ for $k<\tau$.
\end{lemma}

\begin{proof}
A size-6 step has probability $\le C/X_k^2\le C_\varepsilon/n^2$ before $\hat k_1$. On $G$, with $p_k=\beta/W$, $\eta_k$ the centred indicator of a double step and $e_k=H_k-\chi(\hat x_k)$, Taylor's formula and Lemma~\ref{lem:coefloc}(3) give $e_{k+1}=(1-1/\hat x_k)e_k-\eta_k-r_k-\rho_k$ with $|r_k|\le C/\hat x_k$, $|\rho_k|\le C_\varepsilon/n$. With $\hat\rho_k=\prod_{i<k}(1-1/\hat x_i)\in[c_\varepsilon,1]$, $|e_k|\le|\sum_{j<k}\eta_j/\hat\rho_{j+1}|+C_\varepsilon$, and Azuma--Hoeffding bounds the martingale. Note that $\eta_k$ is centred for the full chain, size-6 steps included, since $\mathbb P(\xi_k=1\mid\mathcal F_k)=\beta/W$ in all cases; so $\sum_j\eta_j/\hat\rho_{j+1}$ is a martingale for the full chain, while the recursion for $e_k$ is only used on $G$. On $G\cap E$ and while $X_k\ge x^*/2$, $H_k<M_{X_k}$, so the chain stops only through $H\le0$; the slope bounds on $\chi$ locate $X_\tau$, and~\eqref{eq:chi} gives the last claim.
\end{proof}

\paragraph{Proof of Theorem~\ref{thm:locale}.}
We may assume $n\ge n_\varepsilon$ and $d_0\ge1$: for $d_0=0$ both statements are exact, and finitely many sizes only affect the constants, since $\mathsf N_n,\mathsf F_n\le n$. By Lemma~\ref{lem:FK} and Lemma~\ref{lem:coefloc}(4), $Z^{\mathsf N}\le Cn^{1/3}$, hence $\mathsf N_s\le Cs^{1/3}$; then $\mathsf N_s(d-2)/s\le Cs^{-2/3}$ and $Z^{\mathsf F}\le Cn^{1/3}$ as well, so $\mathbb E[Z^\bullet;(G\cap E)^c]=O_\varepsilon(n^{-2/3})$. On $G\cap E$ the visited sizes are $n,n-3,\dots,X_\tau\ge c_\varepsilon n$, and Lemma~\ref{lem:coefloc}(2) gives
\begin{equation}\label{eq:poids}
\mathcal W^{\mathsf N}_k=(n/X_k)^{1/3}(1+O(1/X_k)),\qquad\mathcal W^{\mathsf F}_k=(X_k/n)^{1/3}(1+O(1/X_k)).
\end{equation}
\emph{Nodes.} Sources contribute $O_\varepsilon(1/n)$, a final value $H_\tau=-1$ has probability $O_\varepsilon(1/n)$, and $\mathcal W^{\mathsf N}_\tau=(n/x^*)^{1/3}(1+O_\varepsilon(\frac{\log n}{\sqrt n}))=(1-2\delta)^{-1/2}+O_\varepsilon(\frac{\log n}{\sqrt n})$.

\emph{Leaves.} The final value vanishes, and the secondary sources $\Sigma^{\mathsf F}_x(d)-\pi\mathsf N_s(d-2)/s$ contribute $O_\varepsilon(1/n)$, exactly as the sources do for nodes. It remains
\[
\sum_{k<\tau}\mathcal W^{\mathsf F}_k\pi\frac{\mathsf N_{X_k-3}(H_k-2)}{X_k-3}.
\]
Since $H_k$ decreases by at least one at every step, the state $H_k=1$ can occur for at most one $k<\tau$. At that step $\mathsf N_{X_k-3}(-1)=0$; replacing this value by the bulk approximation below would contribute only $O_\varepsilon(1/n)$, because $X_k\ge c_\varepsilon n$ and $\mathcal W^{\mathsf F}_k=O(1)$ on $G\cap E$. We therefore apply the node estimate only when $H_k\ge2$, so that $H_k-2\ge0$. Since $H_k/X_k\le\frac12-\frac\varepsilon2$, we have $H_k-2\le(\frac12-\frac\varepsilon2)(X_k-3)$, and $1-2\frac{H_k-2}{X_k-3}=1-2\frac{H_k}{X_k}+O(1/n)$ stays above $\varepsilon$. Hence the node estimate (with $\varepsilon/2$) gives
\[
\mathsf N_{X_k-3}(H_k-2)
=(1-2H_k/X_k)^{-1/2}+O_\varepsilon\Bigl(\frac{\log n}{\sqrt n}\Bigr),
\]
and on $E$, by~\eqref{eq:chi},
\[
(1-2H_k/X_k)^{-1/2}
=(1-2\delta)^{-1/2}(X_k/n)^{1/3}
+O_\varepsilon\Bigl(\frac{\log n}{\sqrt n}\Bigr).
\]
Using $\mathcal W^{\mathsf F}_k=(X_k/n)^{1/3}(1+O(1/X_k))$, $\pi=1+O(1/X_k)$ and $X_k-3=X_k(1+O(1/X_k))$, each summand with $H_k\ge2$ is therefore
\[
(1-2\delta)^{-1/2}X_k^{-1/3}n^{-2/3}
+O_\varepsilon(n^{-3/2}\log n).
\]
The possible single index with $H_k=1$ changes the comparison with this approximate sum by only $O_\varepsilon(1/n)$. The mesh of the visited sizes is $3$, and $|X_\tau-x^*|=O_\varepsilon(\sqrt n\log n)$ on $G\cap E$. Thus the Riemann-sum estimate gives
\[
\begin{aligned}
\sum_{k<\tau}X_k^{-1/3}n^{-2/3}
&=\frac13\int_{x^*}^n x^{-1/3}n^{-2/3}\,dx
  +O_\varepsilon\Bigl(\frac{\log n}{\sqrt n}\Bigr)\\
&=\frac12\bigl(1-(x^*/n)^{2/3}\bigr)
  +O_\varepsilon\Bigl(\frac{\log n}{\sqrt n}\Bigr)\\
&=\delta+O_\varepsilon\Bigl(\frac{\log n}{\sqrt n}\Bigr).
\end{aligned}
\]
Consequently,
\[
\mathsf F_n(d_0)
=\frac{\delta}{\sqrt{1-2\delta}}
 +O_\varepsilon\Bigl(\frac{\log n}{\sqrt n}\Bigr),
\]
which is the asserted leaf profile. \hfill$\square$

\section{The edge window}
\label{sec:bord}

\subsection{Results and discussion}
\label{sec:deuxechelles}

At scale $n$ nothing happens beyond $\frac12$: $K_n/n,D_n/n\le\frac23$ and the limit laws give no mass to $]\frac12,\frac23]$. But the limit densities diverge at $\frac12$ and Theorem~\ref{thm:locale} is not uniform up to $n/2$. The relevant function is
\[
g(y)=\mathbb E\bigl[(-2V)^{-1/2};V<0\bigr],\quad V\sim\mathcal N(y,\tfrac14);\qquad g(y)=\tfrac1{\sqrt2}e^{-y^2}D_{-1/2}(2y),
\]
the second form following from $\int_0^\infty t^{-1/2}e^{-t^2/2-zt}dt=\sqrt\pi\,e^{z^2/4}D_{-1/2}(z)$.

\begin{lemma}
\label{lem:g}
$g$ is smooth with bounded derivatives; $|g(y)-(-2y)^{-1/2}|\le C|y|^{-5/2}$ for $y\le-1$; $g(y)\le g(0)e^{-2y^2}$ for $y\ge0$; and
\[
c_{\mathsf N}:=\int_0^\infty g(y)\,dy=\tfrac12\,\mathbb E|Z|^{1/2}=\frac{\Gamma(3/4)}{2^{3/4}\sqrt\pi}\approx0.41109,\qquad Z\sim\mathcal N(0,1).
\]
\end{lemma}

\begin{proof}
Smoothness follows from Lemma~\ref{lem:phi} below, since $g(y)=\varphi_-(2y)$. For $y\le-1$, write $a=-2y$ and $g(y)=\mathbb E[(a-Z)^{-1/2};Z<a]$; on $|Z|\le a/2$, $(a-Z)^{-1/2}=a^{-1/2}(1+\frac Z{2a}+O(Z^2/a^2))$, the odd term averages to $0$, and the rest is $O(e^{-a^2/16})$. For $y\ge0$ and $v\le0$, $(v-y)^2\ge v^2+y^2$ in the Gaussian density. Finally, with $V=y+\frac12Z$ and Fubini, $\int_0^\infty g=\mathbb E\bigl[\int_0^{|Z|/2}(|Z|-2y)^{-1/2}dy;Z<0\bigr]=\mathbb E[|Z|^{1/2};Z<0]$.
\end{proof}

\begin{theorem}[Uniform approximation of the node profile]
\label{thm:A}
For every admissible size $n\equiv2\pmod3$ and every $0\le d\le M_n$, with $y=(d-n/2)/\sqrt n$,
\[
\bigl|\mathsf N_n(d)-n^{1/4}g(y)\bigr|\le C\,n^{3/16}(1+|y|)^{-7/8}.
\]
In particular $\sup_d\bigl|n^{3/4}\,\mathbb P(D_n=d)-g\bigl(\frac{d-n/2}{\sqrt n}\bigr)\bigr|\le Cn^{-1/16}$.
\end{theorem}

\begin{theorem}[Uniform approximation of the leaf profile]
\label{thm:B}
For every admissible size $n\equiv2\pmod3$ and every $0\le d\le M_n$, with $y=(d-n/2)/\sqrt n$, $\bigl|\mathsf F_n(d)-\frac dn\,n^{1/4}g(y)\bigr|\le C\,n^{3/16}$. In particular
\[
\sup_d\Bigl|\,n^{3/4}\,\mathbb P(K_n=d)-\frac{3d}n\,g\Bigl(\frac{d-n/2}{\sqrt n}\Bigr)\Bigr|\le Cn^{-1/16},
\]
and $n^{3/4}\,\mathbb P(K_n=d)\to\frac32g(y)$ uniformly for $|d-n/2|\le A\sqrt n$.
\end{theorem}

\begin{theorem}[Mass beyond $n/2$]
\label{thm:masse}
$n^{1/4}\,\mathbb P(D_n>n/2)\to c_{\mathsf N}$ and $n^{1/4}\,\mathbb P(K_n>n/2)\to\frac32c_{\mathsf N}=\frac34\mathbb E|Z|^{1/2}=\frac{3\Gamma(3/4)}{2^{7/4}\sqrt\pi}\approx0.61664$. Given $D_n>n/2$ (resp.\ $K_n>n/2$), $(D_n-n/2)/\sqrt n$ (resp.\ $(K_n-n/2)/\sqrt n$) converges in law to the density $g/c_{\mathsf N}$ on $]0,\infty[$.
\end{theorem}

\begin{remark}
\label{rem:raccord}
By Lemma~\ref{lem:g}, for $d\le(\frac12-\varepsilon)n$ one has $n^{1/4}g(y)=(1-2d/n)^{-1/2}+O_\varepsilon(1/n)$, so Theorem~\ref{thm:A} contains the node part of Theorem~\ref{thm:locale}, with error $O_\varepsilon(n^{-1/4})$; it is also uniform over the intermediate zone $A\sqrt n\le n/2-d\le\varepsilon n$. Theorem~\ref{thm:B} is only informative where $\mathsf F_n\gg n^{3/16}$, i.e.\ near $n/2$; in the bulk, Theorem~\ref{thm:locale} is needed. Figure~\ref{fig:num} (right) suggests that both suprema are in fact $O(n^{-1/2})$.
\end{remark}

\begin{figure}[htbp]
\centering
\includegraphics[width=\textwidth]{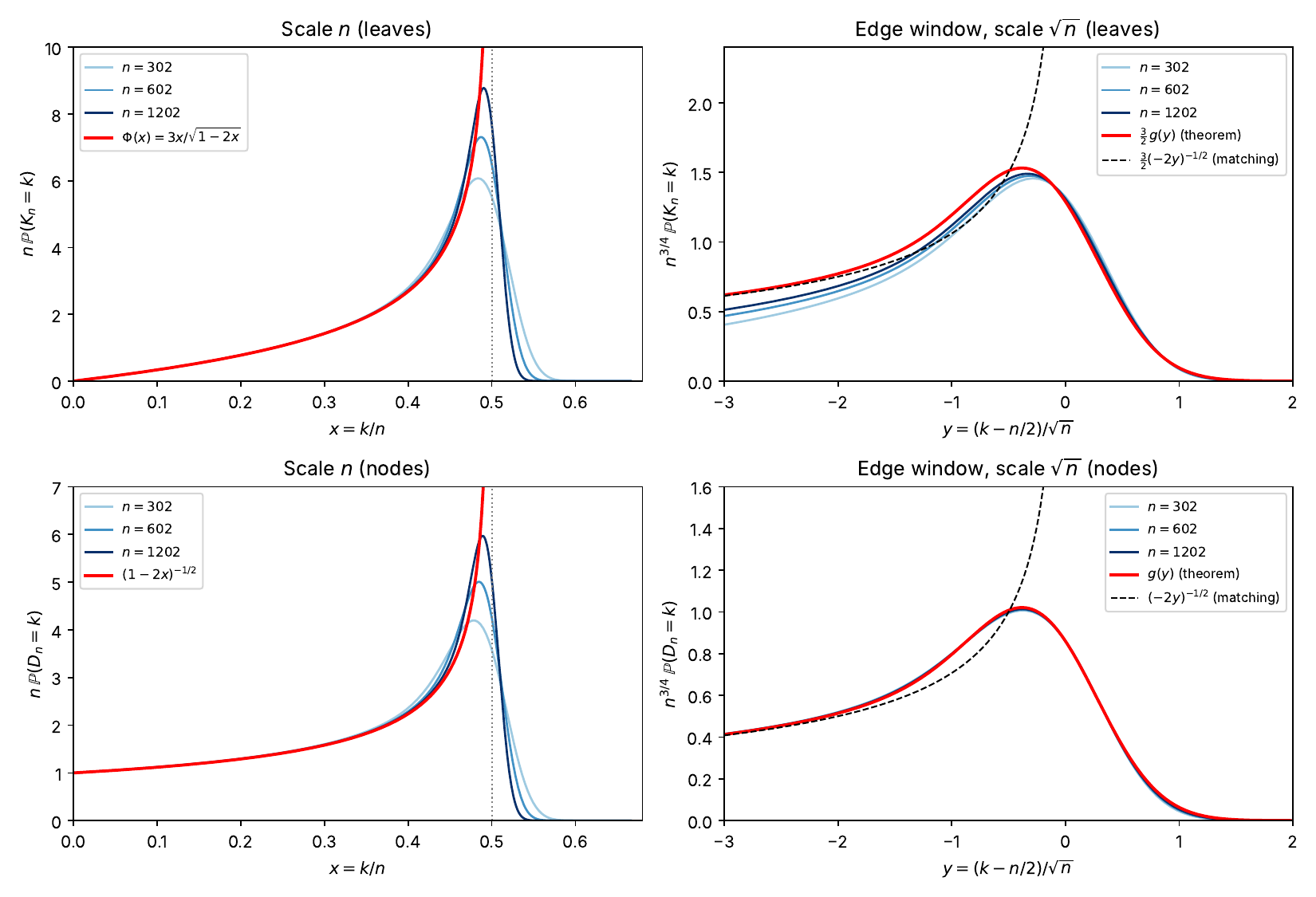}
\caption{Exact mean profiles for $n=302,602,1202$; top: leaves, bottom: nodes. Left: scale $n$, compared with the limit densities of Theorem~\ref{thm:loi}; the peak, whose height grows like $n^{1/4}$, is the edge window. Right: edge window $y=(k-n/2)/\sqrt n$ with ordinate multiplied by $n^{3/4}$, compared with $\frac32g$ and $g$ (Theorems~\ref{thm:B} and~\ref{thm:A}); dashed: the bulk asymptote $(-2y)^{-1/2}$ (times $\frac32$ for leaves).}
\label{fig:deuxregimes}
\end{figure}

\paragraph{Two scalings, one formula.}
Theorems~\ref{thm:loi} and~\ref{thm:A} describe the same object at two scales that no single rescaling reconciles (Figure~\ref{fig:deuxregimes}). At scale $n$, the limit $(1-2x)^{-1/2}$ is a probability density: it carries all the mass, but it is unbounded and gives zero mass to any $o(n)$-neighbourhood of $n/2$. At scale $\sqrt n$, the limit $g$ is bounded and smooth but not a probability density, since $\int_{\mathbb R}g=\infty$; the window $|d-n/2|\le A\sqrt n$ holds a vanishing proportion $\sim n^{-1/4}\int_{-A}^Ag$ of the nodes, that is $\asymp n^{3/4}$ nodes, at a density $n^{1/4}$ times the bulk density $1/n$. Neither limit contains the other. Yet a single explicit formula, $n^{1/4}g\bigl(\sqrt n(\frac dn-\frac12)\bigr)$, is uniformly accurate across both regimes: it is a bulk density sharpened at scale $n^{-1/2}$ near its singularity.

This is unusual. For conditioned Galton--Watson trees, for instance, the profile at scale $\sqrt n$ converges to the local time of the Brownian excursion~\cite{DG97}, a bounded continuous function, and no second scale appears. Here the integrable singularity of the bulk density is the trace of a finer scale whose exponents ($\frac12$ for the width, $\frac14$ for the amplitude) cannot be read off the limit law.

\paragraph{Where the scales come from (heuristic).}
Along the backward chain, $u=(H-X/2)/X^{1/3}$ is nearly a martingale whose remaining variance, down to size $0$, is $X^{1/3}/4$; this is why the heat-equation solutions of Section~\ref{sec:harm} are the right tools. Starting from $d=n/2+y\sqrt n$, $u_0=yn^{1/6}$ fluctuates by $O(n^{1/6})$, i.e.\ $H$ by $O(n^{1/6}X^{1/3})$, which is $\sqrt n$ at $X=n$: this is the width. The chain is absorbed at $H=0$, i.e.\ $u=-X^{2/3}/2$, which for $|u|\asymp n^{1/6}$ happens at $X\asymp n^{1/4}$; the weight $(n/X)^{1/3}\asymp n^{1/4}$ is the amplitude. In the bulk, $u_0\asymp-n^{2/3}$ and absorption happens at $X\asymp n$. For leaves, by contrast, the weights $\mathcal W^{\mathsf F}_k\approx(X_k/n)^{1/3}$ decay, and sizes of order $n$ dominate.

\subsection{Heat-equation tools}
\label{sec:harm}

From now on the chain is the node chain ($\bullet=\mathsf N$). For a state $(x,d)$ let
\[
e=d-x/2,\qquad u=e/x^{1/3},\qquad\sigma(x)=x^{1/3}/4,\qquad p(x,d)=\beta^{\mathsf N}_x(d)/(\alpha^{\mathsf N}_x(d)+\beta^{\mathsf N}_x(d)).
\]

\begin{lemma}
\label{lem:p}
For alive $(x,d)$, $|p(x,d)-d/x|\le4/x$.
\end{lemma}

\begin{proof}
$p=d/(x-2+\epsilon)$ with $\epsilon=2s(1-\pi)w_a/\pi\in[0,3]$ by Lemma~\ref{lem:comb}, and $d\le x$.
\end{proof}

For $\nu\in]0,1[$ let $\psi_\nu(w)=|w|^{-\nu}$ and $\psi^-_\nu(w)=|w|^{-\nu}\mathbf 1_{w<0}$, and for $\psi\in\{\psi_\nu,\psi^-_\nu\}$ let $\varphi(z)=\mathbb E\psi(z+Z)$ and
\[
h(\sigma,u)=\mathbb E\,\psi(u+\sqrt\sigma Z)=\sigma^{-\nu/2}\varphi(u/\sqrt\sigma),
\qquad\text{so that}\qquad\partial_\sigma h=\tfrac12\partial_u^2h .
\]
Note that $g(y)=\varphi_-(2y)$ for $\psi=\psi^-_{1/2}$.

\begin{lemma}
\label{lem:phi}
$|\varphi^{(k)}(z)|\le C(1+|z|)^{-\nu-k}$ for $k\le4$, and for $\psi=\psi_\nu$ also $\varphi(z)\ge c(1+|z|)^{-\nu}$. Hence $|\partial_u^kh|\le C(\sqrt\sigma+|u|)^{-\nu-k}$, and $h\ge c(\sqrt\sigma+|u|)^{-\nu}$ for $\psi=\psi_\nu$.
\end{lemma}

\begin{proof}
Let $\mathsf g$ be the Gaussian density; $\varphi^{(k)}(z)=(-1)^k\int\psi(w)\mathsf g^{(k)}(w-z)dw$ is bounded. For $|z|\ge2$, if the support of $\psi$ stays at distance $\ge|z|/2$ from $z$ (the case of $\psi^-$ with $z\ge2$), the integral is $O(e^{-z^2/32})$. Otherwise, by symmetry take $z\ge2$, $\psi=\psi_\nu$ (or $z\le-2$, $\psi^-_\nu$, identically); with $\chi$ smooth, $0$ on $]-\infty,\frac14]$ and $1$ on $[\frac12,\infty[$, split $\psi=\chi(w/z)w^{-\nu}+(1-\chi(w/z))\psi$. The second part is at distance $\ge z/2$ from $z$, hence contributes $O(e^{-z^2/32})$; for the first, integrate by parts $k$ times, using $|(\chi(w/z)w^{-\nu})^{(k)}|\le C_kw^{-\nu-k}$ on $[z/4,\infty[$. The lower bound is $\varphi(z)\ge(|z|+1)^{-\nu}\mathbb P(|Z|\le1)$.
\end{proof}

\begin{lemma}[One step]
\label{lem:step}
There are $x_0$ and $C$ such that for every alive $(x,d)$ with $x\ge x_0$, writing $(X_1,H_1)$ for the next state of the chain started at $(x,d)$ and $u_1$ for its $u$-coordinate,
\[
\bigl|\mathbb E\,h(\sigma(X_1),u_1)-h(\sigma(x),u)\bigr|\le C\,x^{-3/2}(\sqrt{\sigma(x)}+|u|)^{-\nu},
\]
and $h(\sigma(X_1),u_1)\le C(\sqrt{\sigma(x)}+|u|)^{-\nu}$.
\end{lemma}

\begin{proof}
Write $\sigma=\sigma(x)$, $B_k=(\sqrt\sigma+|u|)^{-\nu-k}$. For each possible next state, $|u_1-u|\le3|u|/x+2(x-6)^{-1/3}$ and $\sigma(X_1)\in[(\frac{x-6}x)^{1/3}\sigma,\sigma]$; for $x_0$ large, $\sqrt{\bar\sigma}+|\bar u|$ lies within a factor $2$ of $\sqrt\sigma+|u|$ on the whole rectangle between $(\sigma,u)$ and the next point, so Lemma~\ref{lem:phi} gives $|\partial_u^kh|\le CB_k$ there; this proves the second claim. A size-6 step has probability $\le C/x^2$ and changes $h$ by $\le CB_0$. Otherwise the chain goes to $(x-3,d-1-\xi)$ with $\xi\sim$ Bernoulli$(p)$. With $\varsigma=(x-3)^{-1/3}$, $\Delta\sigma=\sigma(x)-\sigma(x-3)$ and $r=p-d/x$, so that $\frac12-p=-e/x-r$,
\[
\delta:=u_1-u=e(\varsigma-x^{-1/3})+(\tfrac12-\xi)\varsigma,\quad|\mathbb E\delta|\le C(|u|x^{-2}+x^{-4/3}),\quad|\delta|\le C(|u|x^{-1}+x^{-1/3}),
\]
$\mathrm{Var}\,\delta=p(1-p)\varsigma^2$, and $0\le\varsigma^2-4\Delta\sigma\le Cx^{-5/3}$. Taylor's formula (first in $\sigma$, then in $u$) and $h_\sigma=\frac12h_{uu}$, $h_{\sigma u}=\frac12\partial_u^3h$, $h_{\sigma\sigma}=\frac14\partial_u^4h$ give
\[
\mathbb Eh(\sigma-\Delta\sigma,u+\delta)-h=h_u\mathbb E\delta+\tfrac12h_{uu}(\mathbb E\delta)^2+\tfrac12h_{uu}\bigl(\mathrm{Var}\,\delta-\Delta\sigma\bigr)+\mathcal R,
\]
with $|\mathcal R|\le CB_3|\delta|x^{-2/3}+CB_4x^{-4/3}+CB_3|\delta|^3\le CB_0x^{-3/2}$, using $\sqrt\sigma\asymp x^{1/6}$. Next, $|h_u\mathbb E\delta|+|h_{uu}|(\mathbb E\delta)^2\le CB_0x^{-3/2}$. Finally $\mathrm{Var}\,\delta-\Delta\sigma=-(p-\frac12)^2\varsigma^2+\frac14(\varsigma^2-4\Delta\sigma)$ with $(p-\frac12)^2\le2u^2x^{-4/3}+32x^{-2}$ (Lemma~\ref{lem:p}), so $|h_{uu}|\,|\mathrm{Var}\,\delta-\Delta\sigma|\le CB_2(u^2x^{-2}+x^{-5/3})\le CB_0x^{-2}$.
\end{proof}

\begin{theorem}[Negative moments of the stopping size]
\label{thm:momneg}
For $\nu\in]0,1[$ there is $C_\nu$ such that for all $n\ge17$ and $1\le d_0\le M_n$, with $y=(d_0-n/2)/\sqrt n$,
\[
\mathbb E\bigl[X_\tau^{-2\nu/3}\bigr]\le C_\nu\,n^{-\nu/6}(1+|y|)^{-\nu}.
\]
\end{theorem}

\begin{proof}
Take $\psi=\psi_\nu$ and $F(x,d)=h(\sigma(x),u)$. By Lemmas~\ref{lem:phi} and~\ref{lem:step}, $\mathbb E[F(X_1,H_1)]\le(1+Cx^{-3/2})F(x,d)$ for alive $(x,d)$ with $x\ge x_0$, and $\le CF(x,d)$ for the finitely many alive states with $x<x_0$. As visited sizes are distinct, $F(X_{k\wedge\tau},H_{k\wedge\tau})$ divided by a product bounded by $\Pi_*<\infty$ is a supermartingale, so $\mathbb E[F(X_\tau,H_\tau)]\le\Pi_*F(n,d_0)\le C(n^{1/6}+|y|n^{1/6})^{-\nu}$. At the stopping time, $-1\le H_\tau\le M_{X_\tau}+3$ and $X_\tau\ge11$, so $|u_\tau|\le X_\tau^{2/3}$ and $F(X_\tau,H_\tau)\ge cX_\tau^{-2\nu/3}$.
\end{proof}

\subsection{Proofs of the edge results}

\paragraph{Proof of Theorem~\ref{thm:A}.}
Let $\mathsf G_x(d)=x^{1/4}g\bigl((d-x/2)/\sqrt x\bigr)$ for all $x\ge2$ and integers $d$. With $\psi=\psi^-_{1/2}$ one checks that $\mathsf G_x(d)=2^{-1/2}x^{1/3}h(\sigma(x),u)$: indeed $u+\sqrt{\sigma}Z=x^{1/6}(y+\frac12Z)$. For alive $(x,d)$ define the defect
\[
\mathcal D(x,d)=\mathsf G_x(d)-\bigl[\alpha^{\mathsf N}_x(d)\mathsf G_{x-3}(d-1)+\beta^{\mathsf N}_x(d)\mathsf G_{x-3}(d-2)+\gamma_x\mathsf G_{x-6}(d-1)\bigr].
\]

\begin{lemma}
\label{lem:defect}
$|\mathcal D(x,d)|\le Cx^{-5/4}$ for every alive $(x,d)$.
\end{lemma}

\begin{proof}
There are finitely many alive states with $x<x_0$. For $x\ge x_0$, the bracket equals $W^{\mathsf N}_x\,\mathbb E[\mathsf G(X_1,H_1)]$. Since $W^{\mathsf N}_x(x-3)^{1/3}=x^{1/3}(1+O(x^{-2}))$ and $W^{\mathsf N}_x(x-6)^{1/3}=x^{1/3}(1+O(x^{-1}))$, the latter on an event of probability $O(x^{-2})$, Lemma~\ref{lem:step} gives $|\mathcal D|\le Cx^{1/3}\bigl(x^{-3/2}+x^{-2}\bigr)(\sqrt{\sigma(x)})^{-1/2}\le Cx^{-5/4}$.
\end{proof}

Iterating the definition of $\mathcal D$ along the chain (as in Lemma~\ref{lem:FK}) gives
\[
\mathsf G_n(d_0)=\mathbb E\Bigl[\sum_{k<\tau}\mathcal W_k\mathcal D(X_k,H_k)+\mathcal W_\tau\mathsf G(X_\tau,H_\tau)\Bigr].
\]
Subtracting from Lemma~\ref{lem:FK},
\[
\mathsf N_n(d_0)-\mathsf G_n(d_0)=\mathbb E\Bigl[\sum_{k<\tau}\mathcal W_k\bigl(\Sigma^{\mathsf N}_{X_k}(H_k)-\mathcal D(X_k,H_k)\bigr)+\mathcal W_\tau\bigl(\mathsf N-\mathsf G\bigr)(X_\tau,H_\tau)\Bigr].
\]
We bound each part using $\mathcal W_k\le C(n/X_k)^{1/3}$ and the fact that visited sizes are distinct and exceed $X_\tau$ before $\tau$.
\begin{itemize}
\item By Lemma~\ref{lem:defect}, $\sum_{k<\tau}\mathcal W_k|\mathcal D_k|\le Cn^{1/3}\sum_{x>X_\tau}x^{-1/3-5/4}\le Cn^{1/3}X_\tau^{-7/12}$.
\item By Lemma~\ref{lem:coefloc}(4), and since $H\le2$ occurs at most twice, $\sum_{k<\tau}\mathcal W_k\Sigma^{\mathsf N}_k\le Cn^{1/3}X_\tau^{-4/3}$.
\item Final term. If $H_\tau=0$ and $X_\tau\ge17$: $\mathsf N=1$ and, by Lemma~\ref{lem:g}, $\mathsf G=X_\tau^{1/4}g(-\sqrt{X_\tau}/2)=1+O(X_\tau^{-1})$. If $H_\tau>M_{X_\tau}$: $\mathsf N=0$ and $\mathsf G\le CX_\tau^{1/4}e^{-X_\tau/18}$. If $X_\tau<17$: $|\mathsf N-\mathsf G|\le C$. If $H_\tau=-1$: $|\mathsf N-\mathsf G|\le C$, but this requires a double step from $H=1$, of probability $\beta^{\mathsf N}_x(1)/W^{\mathsf N}_x\le C/x$, whence $\mathbb E[\mathcal W_\tau;H_\tau=-1]\le Cn^{1/3}\mathbb E[X_\tau^{-4/3}]$.
\end{itemize}
Altogether, since $X_\tau\ge11$, $|\mathsf N_n(d_0)-\mathsf G_n(d_0)|\le Cn^{1/3}\mathbb E[X_\tau^{-7/12}]$, and Theorem~\ref{thm:momneg} with $\nu=\frac78$ gives $Cn^{1/3-7/48}(1+|y|)^{-7/8}=Cn^{3/16}(1+|y|)^{-7/8}$. The case $d_0=0$ follows from Lemma~\ref{lem:g}. Finally $n^{3/4}\mathbb P(D_n=d)=n^{-1/4}\mathsf N_n(d)$. \hfill$\square$

\paragraph{Proof of Theorem~\ref{thm:B}.}
Let $\mathsf Q_x(d)=\frac dx\mathsf G_x(d)$ and, for alive $(x,d)$, with $s=x-3$,
\[
\mathcal D^{\mathsf F}(x,d)=\mathsf Q_x(d)-\bigl[\alpha^{\mathsf F}_x(d)\mathsf Q_s(d-1)+\beta^{\mathsf F}_x(d)\mathsf Q_s(d-2)+\gamma_x\mathsf Q_{x-6}(d-1)\bigr]-\pi\frac{\mathsf G_s(d-2)}s .
\]

\begin{lemma}
\label{lem:defectF}
$|\mathcal D^{\mathsf F}(x,d)|\le Cx^{-5/4}$ for every alive $(x,d)$.
\end{lemma}

\begin{proof}
Write $A_1=\mathsf G_s(d-1)$, $A_2=\mathsf G_s(d-2)$, $A_6=\mathsf G_{x-6}(d-1)$, $\delta=d/x$, and substitute $\mathsf Q_x(d)=\delta\bigl[\alpha^{\mathsf N}_xA_1+\beta^{\mathsf N}_xA_2+\gamma_xA_6+\mathcal D(x,d)\bigr]$. A direct computation with the explicit coefficients gives
\[
\mathcal D^{\mathsf F}=\delta\,\mathcal D(x,d)+\frac\pi s\,(1-3\delta+3\delta^2)\,(A_1-A_2)+O(x^{-2})(|A_1|+|A_2|)+O(x^{-3})|A_6| .
\]
Indeed the coefficient of $A_1$ is $\frac\pi s[(s-d+1)\delta-(s-d)\frac{d-1}s]+O(x^{-2})=\frac\pi s(1-3\delta+3\delta^2)+O(x^{-2})$, that of $A_2$ is $\frac\pi s[d\delta-\frac{(d-1)(d-2)}s-1]=-\frac\pi s(1-3\delta+3\delta^2)+O(x^{-2})$, and that of $A_6$ is $\gamma_x(\delta-\frac{d-1}{x-6})=O(x^{-3})$. Now $|\mathsf G_s|\le Cx^{1/4}$, $|A_1-A_2|\le s^{1/4}\|g'\|_\infty s^{-1/2}$, and Lemma~\ref{lem:defect} applies.
\end{proof}

As for nodes, the leaf chain gives
\[
\mathsf F_n(d_0)-\mathsf Q_n(d_0)=\mathbb E\Bigl[\sum_{k<\tau}\mathcal W^{\mathsf F}_k\Bigl(\pi\frac{(\mathsf N_s-\mathsf G_s)(H_k-2)}{s}+R_k-\mathcal D^{\mathsf F}_k\Bigr)+\mathcal W^{\mathsf F}_\tau(\mathsf F-\mathsf Q)(X_\tau,H_\tau)\Bigr],
\]
where $s=X_k-3$ and $0\le R_k\le C(X_k^{-1}\mathbf 1_{H_k\le2}+X_k^{-2})$ (Lemma~\ref{lem:coefloc}(4)). By Theorem~\ref{thm:A}, $|\mathsf N_s(d')-\mathsf G_s(d')|\le Cs^{3/16}$ for all $d'$ (for $d'\notin[0,M_s]$, $\mathsf N_s=0$ and $\mathsf G_s\le C$). The final term is bounded: $\mathsf F$ and $\mathsf Q$ both vanish at $H_\tau=0$, and $|\mathsf Q|\le C$ at the other stopped states. Since $\mathcal W^{\mathsf F}_k\le C$ and visited sizes are distinct,
\[
|\mathsf F_n(d_0)-\mathsf Q_n(d_0)|\le C\sum_{x\le n}\bigl(x^{-13/16}+x^{-5/4}+x^{-2}\bigr)+C\le Cn^{3/16}.
\]
Finally $n^{3/4}\mathbb P(K_n=d)=\frac{3n}{n+1}n^{-1/4}\mathsf F_n(d)$ and $\frac{3d}n=\frac32+O(A/\sqrt n)$ in the window. \hfill$\square$

\paragraph{Proof of Theorem~\ref{thm:masse}.}
By Theorems~\ref{thm:A} and~\ref{thm:B}, Riemann sums of mesh $n^{-1/2}$ give, for every $A>0$,
\[
n^{1/4}\mathbb P\bigl(n/2<D_n\le n/2+A\sqrt n\bigr)\to\int_0^Ag,\qquad n^{1/4}\mathbb P\bigl(n/2<K_n\le n/2+A\sqrt n\bigr)\to\tfrac32\int_0^Ag .
\]
Since every leaf is a node, $\mathsf F_n(d)\le\mathsf N_n(d)$, hence $\mathbb P(K_n>m)\le3\,\mathbb P(D_n>m)$. Everything therefore reduces to the node tail bound
\begin{equation}\label{eq:queue}
\lim_{A\to\infty}\limsup_{n\to\infty}n^{1/4}\,\mathbb P\bigl(D_n>n/2+A\sqrt n\bigr)=0,
\end{equation}
and the conditional statements follow by taking ratios. For the node chain from an alive $(x_0,d_0)$ with $d_0\ge x_0/2$, let $\check V_k=(H_k-X_k/2)\,x_0^{-1/6}X_k^{-1/3}$, so that $\check V_0=y:=(d_0-x_0/2)/\sqrt{x_0}$.

\begin{lemma}
\label{lem:expo}
$\mathbb P(\exists k\le\tau:\check V_k\le v)\le Ce^{-(y-v)^2}$ for $0\le v\le y$.
\end{lemma}

\begin{proof}
Let $(x,d)$ be alive with $\check V\ge0$ and $\theta\ge0$. On a size-3 step, $\check V'=c_x\check V-(\eta+r)s_x$, with $s_x=x_0^{-1/6}(x-3)^{-1/3}$, $\eta=\xi-p$ centred with range $1$, $|r|\le4/x$ (Lemma~\ref{lem:p}) and $c_x=(1-\frac1x)(1-\frac3x)^{-1/3}\ge(1-\frac1x)(1+\frac1x+\frac2{x^2})\ge1$, because $(1-t)^{-1/3}$ has positive Taylor coefficients. On a size-6 step, $e$ becomes $e+2$, so $\check V'\ge\check V$. By Hoeffding's lemma, $\mathbb E[e^{-\theta\check V'}]\le e^{-\theta\check V}K_x$ with $K_x=\exp(4\theta s_x/x+\theta^2s_x^2/8)$. Over distinct sizes, $\sum s_x^2\le1+x_0^{-1/3}$ and $\sum s_x/x\le Cx_0^{-1/6}$. Stopping at $T=\min\{k:\check V_k\le v\}\wedge\tau$ yields $e^{-\theta v}\,\mathbb P(\exists k\le\tau:\check V_k\le v)\le e^{-\theta y}\exp\bigl(C\theta x_0^{-1/6}+\theta^2(1+x_0^{-1/3})/8\bigr)$; take $\theta=4(y-v)/(1+x_0^{-1/3})$ and use $x_0^{-1/3}\le\frac12$.
\end{proof}

\begin{lemma}
\label{lem:queueN}
Let $\nu=\frac{15}{16}$ and $c_\nu=1-\frac1{2\nu}$. For $x\ge17$, $x/2\le d\le M_x$ and $y=(d-x/2)/\sqrt x$,
\[
\mathsf N_x(d)\le Cx^{1/4}\bigl[x^{-(2\nu-1)/12}(1+y)^{-\nu}+e^{-c_\nu y^2}\bigr].
\]
\end{lemma}

\begin{proof}
Let $B=\{\exists k<\tau:H_k\le2\}$. The sources $Cx^{-2}$ contribute $\le Cx^{1/3}X_\tau^{-4/3}$; the sources $Cx^{-1}\mathbf 1_{d\le2}$ and the final value at $H_\tau=0$ (which forces $H_{\tau-1}\le2$) contribute $\le C(x/X_\tau)^{1/3}\mathbf 1_B$; the final value when $X_\tau<17$ is $\le Cx^{1/3}$. Hence $Z^{\mathsf N}\le Cx^{1/3}X_\tau^{-2\nu/3}+C(x/X_\tau)^{1/3}\mathbf 1_B$. Theorem~\ref{thm:momneg} bounds the first term, and Hölder's inequality with exponent $2\nu$ bounds the second by $Cx^{1/4}(1+y)^{-1/2}\mathbb P(B)^{c_\nu}$. On $B$, $H_k-X_k/2<0$ for some $k<\tau$, so Lemma~\ref{lem:expo} with $v=0$ gives $\mathbb P(B)\le Ce^{-y^2}$.
\end{proof}

At most $2\sqrt n$ depths have $y\in[j,j+1[$, and $y\le\sqrt n$. By Lemma~\ref{lem:queueN}, for $A\ge1$,
\[
n^{1/4}\mathbb P(D_n>n/2+A\sqrt n)=n^{-3/4}\!\!\sum_{d>n/2+A\sqrt n}\!\!\mathsf N_n(d)\le Cn^{-\frac{2\nu-1}{12}}\sum_{j\le\sqrt n}(1+j)^{-\nu}+C\sum_{j\ge\lfloor A\rfloor}e^{-c_\nu j^2}.
\]
The first term is $O(n^{(7-8\nu)/12})=O(n^{-1/24})$ and the second tends to $0$ as $A\to\infty$, proving~\eqref{eq:queue}. \hfill$\square$

\section{Conclusion}
\label{sec:conclusion}

The depth profile of a uniform closed linear $\lambda$-term is governed by a linear scale, where the depth of a leaf follows a $\Beta(2,\frac12)$ law after normalisation, and by a finer scale around $n/2$, of width $\sqrt n$ and relative amplitude $n^{1/4}$, where the singularity of the limit density is resolved by a parabolic cylinder function. The grafting identity reduces the uniform model to a growth process with rare perturbations, and the backward chain turns profile questions into questions about a Markov chain with weights; in the edge window, solutions of the heat equation evaluated along this chain replace the concentration arguments that suffice in the bulk.

Several questions remain open. The proved rate $n^{-1/16}$ in the uniform approximations is far from the rate $n^{-1/2}$ suggested by the exact data (Figure~\ref{fig:num}), and a uniform approximation of the leaf profile with small relative error in the bulk, which would contain Theorem~\ref{thm:locale} for leaves, is still missing. The \emph{height}, i.e.\ the maximal depth, is not covered by our mean-profile results. The mean number of nodes at depth $n/2+y\sqrt n$ is about $n^{1/4}g(y)$, with $g(y)\le g(0)e^{-2y^2}$; if this persisted for $y$ of order $\sqrt{\log n}$, the height would be $n/2+O(\sqrt{n\log n})$, but this lies beyond our error terms and would also require second-moment information on the profile. Joint laws of several depths, and the correction terms in $1/n$ of the moments, are also natural targets. Finally, it would be interesting to know whether the two-scale phenomenon persists for other families of $\lambda$-terms (general, affine, planar) and for the corresponding classes of maps through the bijections of~\cite{Z16}.

\end{document}